\documentclass{article}

\usepackage{icrctc07}
\pdfoutput=1

\title{The Highest Energy Neutrinos}
\shorttitle{The Highest Energy Neutrinos}

\authors{Francis Halzen}
\shortauthors{Francis Halzen}

\afiliations{Department of Physics, University of Wisconsin,
Madison, WI 53706, USA}

\email{E-mail: halzen@icecube.wisc.edu}

\abstract{Measurements of the arrival directions of cosmic rays have not revealed their sources. High energy neutrino telescopes attempt to resolve the problem by detecting neutrinos whose directions are not scrambled by magnetic fields. The key issue is whether the neutrino flux produced in cosmic ray accelerators is detectable. It is believed that the answer is affirmative, both for the galactic and extragalactic sources, provided the detector has kilometer-scale dimensions. We revisit the case for kilometer-scale neutrino detectors in a model-independent way by focussing on the energetics of the sources. The real breakthrough though has not been on the theory but on the technology front: the considerable technical hurdles to build such detectors have been overcome.

Where extragalactic cosmic rays are concerned an alternative method to probe the accelerators consists in studying the arrival directions of neutrinos produced in interactions with the microwave background near the source, i.e. within a GZK radius. Their flux is calculable within large ambiguities but, in any case, low. It is therefore likely that detectors that are larger yet by several orders of magnitudes are required. These exploit novel techniques, such as detecting the secondary radiation at radio wavelengths emitted by neutrino induced showers.}

\begin{document}
 
\maketitle

\section{Cosmic Rays and Neutrinos}

\begin{figure*}[t]
 \centering
\includegraphics[width=5.5in]{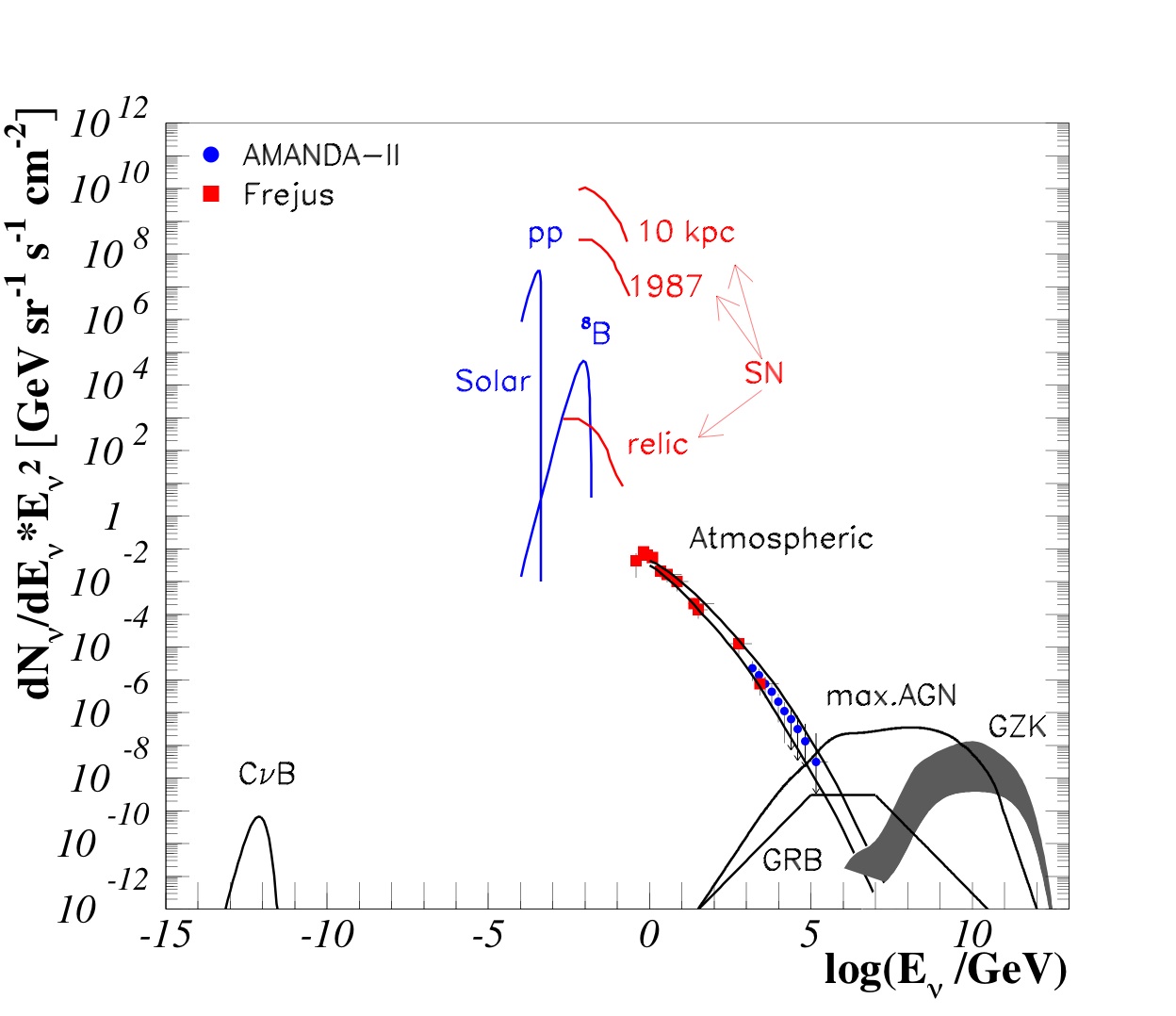}
 
 \caption{The neutrino sky from the lowest energy neutrinos produced in the big bang to the highest energies associated with the sources of the cosmic rays, here assumed to be gamma ray bursts or, alternatively, active galaxies. These will be the target of kilometer-scale neutrino detectors such as IceCube and KM3NeT. Neutrinos at intermediate energies, produced in the sun, supernovae and in collisions of cosmic rays in the atmosphere, have been studied by SuperK and similar detectors\protect\cite{reviewslow} \label{neutrinosky}}
\end{figure*}

An illustration of the neutrino sky is shown in Fig.~\ref{neutrinosky} displaying a spectrum ranging from microwave neutrinos produced in the big bang to GZK neutrinos associated with the highest energy cosmic rays\cite{julia}. The GZK neutrinos are the decay products of pions produced in the interaction of cosmic rays with microwave photons. These are the same interactions that shape the Greissen-Zatsepin-Kuzmin absorption feature in the spectrum, hence their name. Prominently displayed in the figure is the flux of the highest energy atmospheric neutrinos observed up to $\sim 100$\,TeV by the AMANDA experiment\cite{muenich}. This beam, very successfully mined for particle physics by the Superkamiokande-generation of experiments, will be exploited at yet higher energies\cite{GHM}. Because of its steep spectrum, events above several hundreds of TeV become very rare, leaving a clear neutrino sky to be explored for sources of cosmic neutrinos beyond the sun. Neutrino telescopes will open some ten orders of magnitude in neutrino wavelength, from their tens of GeV threshold to the EeV energy of GZK neutrinos.\footnotemark \ The existence of cosmic neutrinos with yet higher energy are a matter of speculation; they could be the decay products of cosmic remnants or topological defects associated with phase transitions in the early universe\cite{reviews}.
 \footnotetext{We use units GeV, TeV, PeV and EeV, increasing energy in steps of one thousand.} 
 
 % Fig.1
 
\begin{figure}[t]
 \includegraphics[width=2.75in]{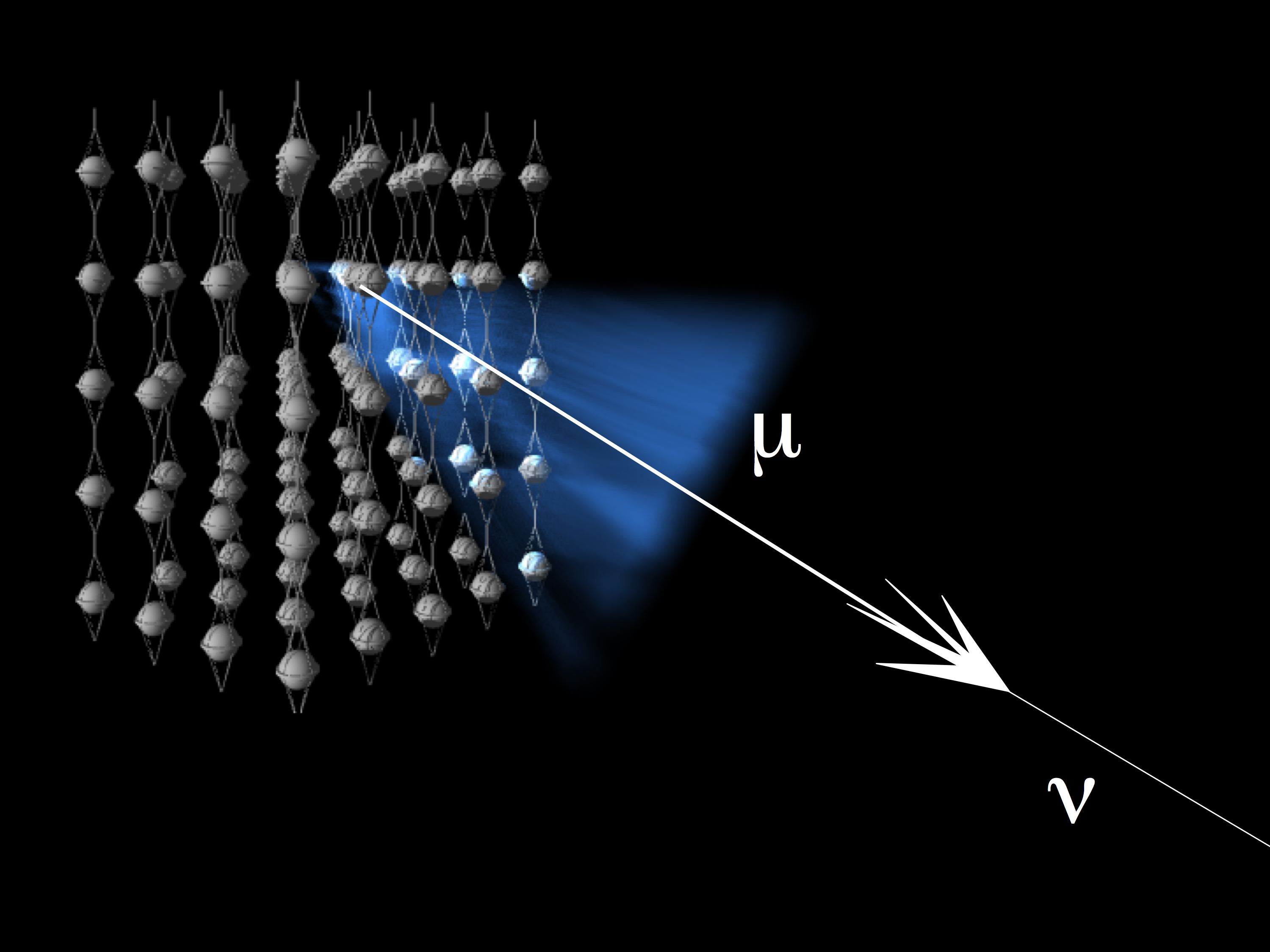}
  
 \caption{Neutrino telescopes take advantage of the large cross section of high-energy neutrinos and the long range of the muons produced. The detector consists of a lattice of photomultipliers deployed in a shielded and optically clear medium that is transformed into a Cherenkov detector. \label{detector}}
 \end{figure}

The neutrino fluxes anticipated\cite{reviews} from non-thermal astronomical sources and, therefore, candidate cosmic ray accelerators such as active galaxies (AGN) and gamma ray bursts (GRB), dominate the atmospheric flux above $\sim 1$\,PeV; see Fig.~\ref{neutrinosky}. The energy fluxes are at the level of the flux, as far as we know unobservable, of big bang neutrinos. The venture can nevertheless succeed by exploiting the relatively large high energy neutrino interaction cross sections in combination with detectors of gigaton size.  Standard model physics is sufficient to establish that the cosmic fluxes shown are observable in a volume of 1 kilometer cubed instrumented with photomultipliers\cite{reviews}. Neutrino telescopes detect the Cherenkov radiation from secondary particles produced in the interactions of high energy neutrinos in highly transparent and well shielded deep water or ice. At the higher energies the neutrino cross section grows and secondary muons travel up to tens of kilometers to reach the detector from interactions outside the instrumented volume; see Fig.~\ref{detector}.
 
 %Fig. 2

The construction of kilometer-scale instruments such as IceCube at the South Pole and the future KM3NeT detector in the Mediterranean, have been made possible by development efforts that resulted in the commissioning of prototypes that are two orders of magnitude smaller, AMANDA and ANTARES\cite{antares}. Their successful technologies have, in turn, relied on pioneering efforts by the DUMAND\cite{dumand} and Baikal\cite{baikal}, as well as the Macro and SuperK collaborations\cite{reviewslow}. While much larger than the latter, kilometer scale neutrino telescopes are insensitive to neutrinos in the MeV-GeV energy range and have a typical threshold of tens of GeV; this is the price one pays for reaching large detection volume. IceCube\cite{ice3} is under construction and taking data with a partial array of 1320 ten inch photomultipliers positioned between 1500 and 2500 meter and deployed as beads on 22 strings below the geographic South Pole. Its effective telescope area already exceeds that of its predecessor AMANDA by roughly one order of magnitude. The detector will grow by another $14{\sim}18$ strings in the 2007-08 Antarctic summer to be completed in 2011 with 80 strings\,\cite{karle}.

The AMANDA experiment has observed neutrinos with energies as high as ${\sim}100$\,TeV, at a rate consistent with the flux of atmospheric neutrinos extrapolated from lower energy measurements; see Fig.~\ref{neutrinosky}. The fluxes of cosmic neutrinos shown in the figure at higher energies are, in contrast, a matter of speculation. It is known that non-thermal sources such as supernova remnants, AGN and GRB accelerate electrons to energies close to 1\,PeV. Their existence is inferred from the observation of TeV gamma rays whose spectrum extends to $\sim 100$\,TeV in some sources\cite{reviewsgamma}. All observations can be accommodated in models where the origin of the photons, from radio to TeV energy, is synchrotron radiation by the electrons and, at the highest energies, inverse Compton scattering of ambient light, primarily the synchrotron photons themselves. There is no reason why non-thermal sources would not accelerate protons or nuclei along with the electrons, turning them into sources of cosmic rays; unfortunately, at this time, no evidence for such cosmic ray accelerators exists. A conclusive signature for the presence of cosmic rays in the sources is the production of pions on ambient radiation and matter. Pion production is revealed by observing the decay products, photons and neutrinos. While it has been a challenge to disentangle such pionic photons from those produced by purely electromagnetic processes\cite{disentangle}, charged pions decaying into neutrinos yield incontrovertible evidence. The anticipated neutrino fluxes are shown in Fig.~\ref{neutrinosky}, assuming that AGN or, alternatively, GRB, happen to be the correct guess for the unknown sources of the cosmic rays. If not, the real sources may be revealed by neutrinos that, unlike charged primaries, point back to their site of origin. Neutrino astronomy must succeed because, after all, cosmic rays exist. The critical question is whether our estimate of the level of the neutrino fluxes associated with the observed cosmic rays is robust; it sets the scale of the detector.

\begin{figure*}[t]
\centering
\includegraphics[width=5in]{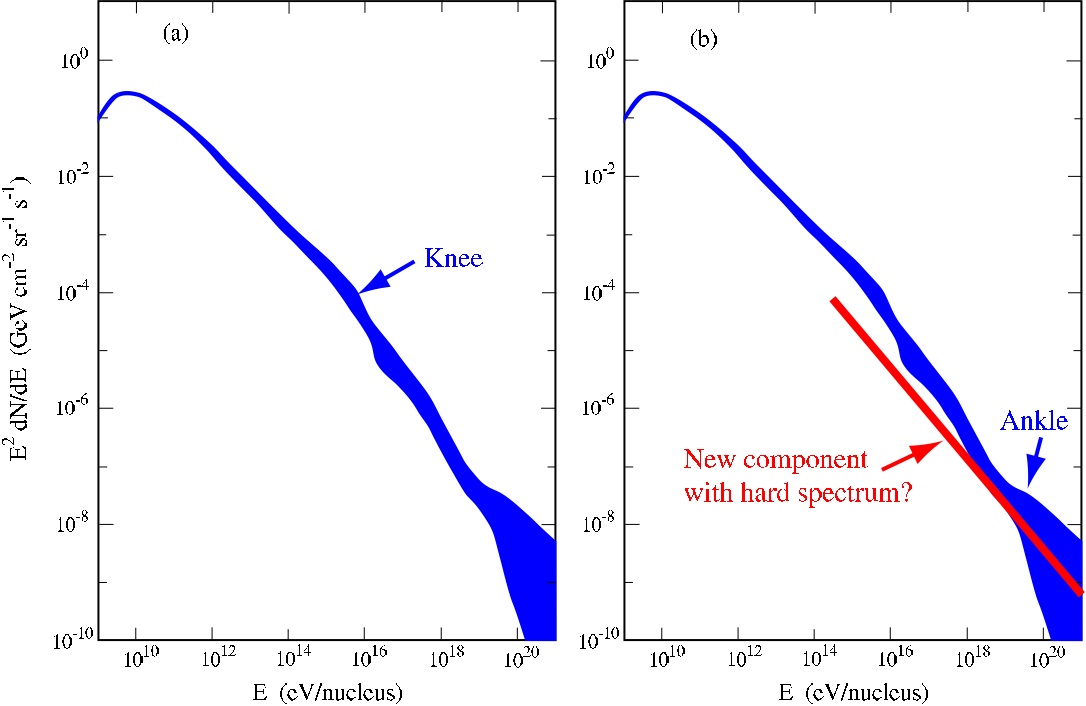}

\caption{At the energies of interest here, the cosmic ray spectrum consists of a sequence of 3 power laws. The first two are separated by the ``knee" (left panel), the second and third by the ``ankle". There is evidence that the cosmic rays beyond the ankle are a new population of particles produced in extragalactic sources; see right panel.\label{gaisserschematic}}

\end{figure*}

Thus the faith of neutrino astronomy is intertwined with cosmic ray physics beyond the traditional subject of GZK neutrinos to which we will return later. While kilometer-scale neutrino detectors are discovery experiments with missions as diverse as particle physics and the search for dark matter -- see Table \ref{nu-telescopes} -- their size as astronomical telescopes is very much anchored to the observed fluxes of galactic and extragalactic cosmic rays. Cosmic accelerators produce particles with energies in excess of $10^8$\,TeV; we still do not know where or how. The flux of cosmic rays observed at Earth follows a broken power law; see Fig.~\ref{gaisserschematic}. The two power laws are separated by a feature dubbed the ``knee". Circumstantial evidence exists that cosmic rays, up to perhaps EeV energy, originate in galactic supernova remnants. Any association with our Galaxy disappears in the vicinity of a second feature in the spectrum referred to as the ``ankle". Above the ankle, the gyroradius of a proton in the galactic magnetic field exceeds the size of the Galaxy and it is generally assumed that we are witnessing the onset of an extragalactic component in the spectrum that extends to energies beyond 100 EeV. Observation of the GZK feature in the HiRes and Auger spectra near the energy threshold for pion production on microwave photons, provides further support for the existence of an extragalactic component.\footnotemark\ \footnotetext{That the cutoff is associated with the upper energy range of the accelerator(s) can at this point not be ruled out.} While the enigmatic nature of the highest energy cosmic rays has been widely advertised, it is also a fact that the origin of the galactic cosmic rays has not been established.

%table 1
\begin{table*}[t]
\caption{Built as discovery instruments, neutrinos telescopes nevertheless target a range of particle and astrophysics problems.\label{nu-telescopes}}
\medskip
\begin{tabular}{l|l}
\multicolumn{1}{c}{Favorite Sources}&\multicolumn{1}{c}{Possible Science}\\
\hline
Atmospheric& Oscillations\\
($\sim 100,000$ per year, up to 1000 TeV, charm?)& New interactions\\
&  Test of relativity and equivalence principle\\[3mm]
GRB& Sources of cosmic rays\\
(successful and failed)& Test of Lorentz invariance\\
& Planck scale physics, quantum decoherence\\[3mm]
AGN& Sources of cosmic rays\\[3mm]
Starburst galaxies\\[3mm]
Supernova remnants& Sources of cosmic rays\\
Also microquasars, magetars, PWNe, binaries\\
unidentified Egret sources, plane of the galaxy\\[3mm]
Cosmic rays ineracting with microwave photons\hspace*{2em}& Identify sources of cosmic rays\\
& Neutrino cross section at EeV energy\\[3mm]
Dark matter& Annihilation in the sun, mostly spin-dependent\\[3mm]
Cosmic rays interacting with the sun& Backrounds to WIMP search\\[3mm]
Supernovae explosion& Deleptonization, TeV emission, hierarchy, $\sin\theta_{13}$\\
\hline
\end{tabular}
\end{table*}

%Fig.3

\section{Neutrinos Associated with Extragalactic Cosmic Rays}

It is routinely emphasized how small cosmic ray particle fluxes are; this may be besides the point. The energetics of the accelerator is likely to be more revealing. A hint in this direction comes from conventional astronomy. While the diffuse universal flux of photons falls by eighteen orders of magnitude between microwave and GeV-energy, the energy carried by the particle flux drops by less than five. Sources are known that emit most of their energy in TeV photons. The energy is the key. The argument has been well advertised for galactic cosmic rays where energetics points at their supernova origin. By integrating the observed flux in Fig.~\ref{gaisserschematic} we can obtain the energy density $\rho_E$ of cosmic rays in the galaxy from the relation that flux${}={}$velocity${}\times{}$density, or
\begin{equation}
4\pi \int  dE \left\{ E{dN\over dE} \right\} =  c\rho_E\,.
\end{equation}
The answer is that $\rho_E \sim 10^{-12}$\,erg\,cm$^{-3}$. This is also the value of the corresponding energy density $B^2/8\pi$ of the microgauss magnetic field in the galaxy. The accelerating power needed to maintain this energy density is $10^{-26}$\,erg/cm$^3$s given that the average containment time of the cosmic rays in our galaxy is $3\times10^6$\,years. For a nominal volume of the galactic disk of $10^{67}$\,cm$^3$ this requires an accelerator delivering $10^{41}$\,erg/s. This happens to be 10\% of the power produced by supernovae releasing $10 ^{51}\,$erg, or the one percent of a solar mass that is not released in neutrinos, every 30 years. This coincidence is the basis for the idea that shocks produced by supernovae expanding into the interstellar medium are the origin of the galactic cosmic rays\cite{ADV94}.

Let's follow the same logic for the extragalactic component in Fig.~\ref{gaisserschematic}. The flux above the ankle is often summarized as ``one $10^{19}$\,eV particle per kilometer square per year per
 steradian". This can be translated into an energy flux
\begin{eqnarray}
E \left\{ E{dN\over dE} \right\}&=& {10^{19}\,{\rm eV} \over \rm (10^{10}\,cm^2)(3\times 10^7\,sec) \, sr}\nonumber \\ 
&=& 3\times 10^{-8}\rm\, GeV\ cm^{-2} \, s^{-1} \, sr^{-1} \,. \nonumber\\ 
&&
\label{eq:flux}\end{eqnarray}
Following the procedure applied above to the galactic component we obtain an energy density of
\[
\rho_E = {4\pi\over c} \int_{E_{\rm min}}^{E_{\rm max}} {3\times 10^{-8}\over E} dE \, {\rm {GeV\over cm^3}} \simeq 10^{-19} \, {\rm {TeV\over cm^3}} \,,
\]
taking the extreme energies of the accelerator(s) to be $E_{\rm max} / E_{\rm min} \simeq 10^3$.

The energy content derived ``professionally" by integrating the spectrum in Fig.~\ref{gaisserschematic} assuming an $E^{-2}$
energy spectrum, typical of shock acceleration, with a GZK cutoff,  is $\sim 3 \times 10^{-19}\rm\,erg\ cm^{-3}$.
This is within a factor of our back-of-the-envelope estimate recalling that 1\,TeV = 1.6\,erg. The energy density represents the universe's filling factor in cosmic rays, equivalent to 410 microwave-energy photons per cubic centimeter.

The power required for a population of sources to generate this energy density over
the Hubble time of $10^{10}$\,years is $\sim 3 \times 10^{37}\rm\,erg\ s^{-1}$ per (Mpc)$^3$ or,
as often quoted in the literature, $\sim 5\times10^{44}\rm\,TeV$ per (Mpc)$^3$ per year.
This works out to\cite{TKG}
\begin{itemize}
\item $\sim 3 \times 10^{39}\rm\,erg\ s^{-1}$ per galaxy,
\item $\sim 3 \times 10^{42}\rm\,erg\ s^{-1}$ per cluster of galaxies,
\item $\sim 2 \times 10^{44}\rm\,erg\ s^{-1}$ per active galaxy, or
\item $\sim 2 \times 10^{51}$\,erg per cosmological gamma ray burst.
\end{itemize}
The coincidence between the last two numbers with the observed output in electromagnetic radiation of these sources, explains why AGN and GRB emerged as leading candidates for the cosmic accelerators. In either case, it suffices that the shocks associated with acceleration near the black hole dump roughly equal energy in electrons and protons to accommodate the observed coincidence, with the electron energy observed as radiation by synchrotron and inverse Compton scattering.

For GRB the argument is reminiscent of the one favoring the ``evidence" for galactic supernova as cosmic ray accelerators. Observations show, within one gigaparsec cubed, 300 GRB dumping about $10 ^{51}\,$erg of energy into the universe in a single year. They therefore supply roughly $10^{44} \rm erg/yr/Mpc^3$ in radiation and, assuming equal energy in protons, we conclude that they represent an environment that can accommodate the observed energetics of the extragalactic cosmic rays. A problem is that the same argument can be made to validate AGN as the sources of the highest energy cosmic rays; see above. In the end, the answer may lay elsewhere. At this point we should emphasize, in either case, the challenge remains to explain the acceleration of particles with energies as high as $10^8$\,TeV, an energy which is in either source near the dimensionally allowed upper limit set by the Hillas formula.

\begin{figure}
\includegraphics[width=2.8in]{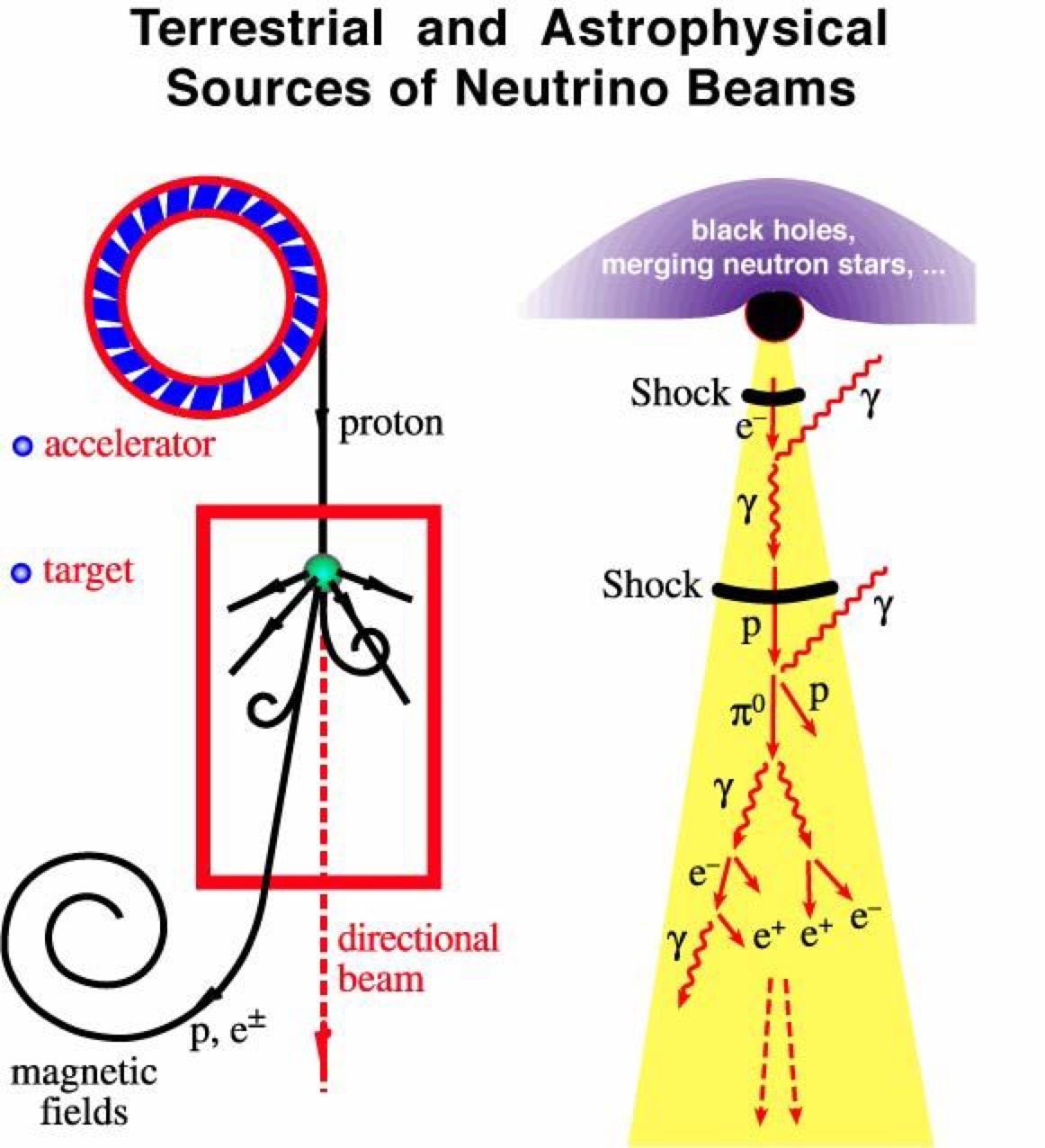}

\caption{Cosmic beam dumps exist: sketch of cosmic ray accelerator producing photons. The charged pions that are inevitably produced along with the neutral pions will decay into neutrinos.\label{nubeams}}
\end{figure}

\begin{figure*}[t]

\includegraphics[width=5.5in]{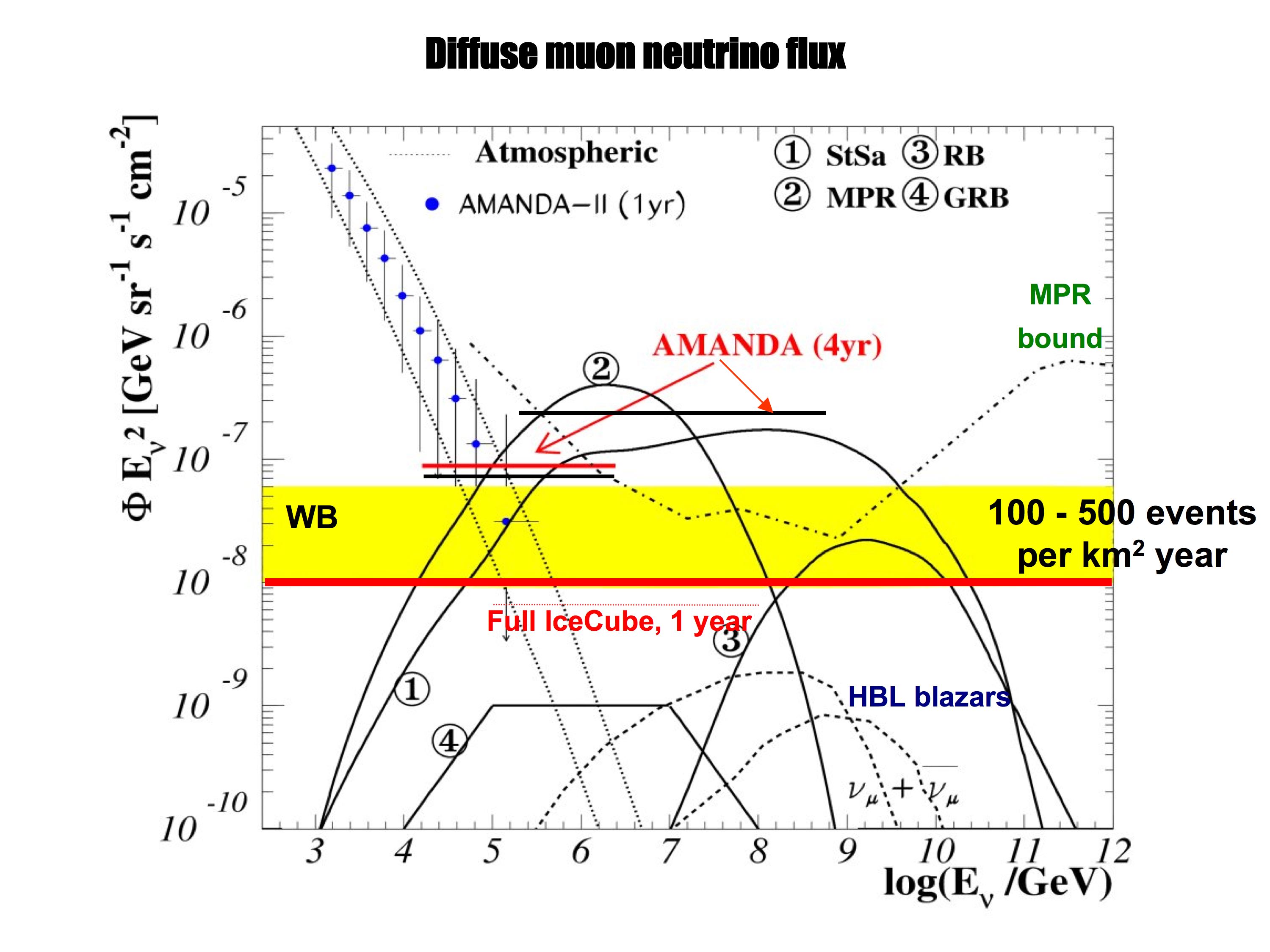}

\caption{Our estimate of the flux of neutrinos associated with the
 sources of the highest energy cosmic rays (the shaded range labeled WB)
 is compared to the limits established by the AMANDA experiment reached with
 800 days of data\cite{hill}. AMANDA's sensitivity is within a factor of 2 of the most optimistic predictions. Also shown are fluxes predicted by specific models
 of cosmic ray accelerators: active galaxies labeled StSa\protect\cite{agn}
 and MPR\protect\cite{MPR}, GRB\protect\cite{guetta} and
 the diffuse flux produced by cosmic ray producing active galaxies on
 microwave photons\protect\cite{RB} labelled RB. Data for the background
 atmospheric neutrino flux are from the AMANDA experiment. The IceCube experiment will be sensitive to all predictions after a few years of operation of the full detector. It has sensitivity to the larger fluxes by operating the partially completed detector that already now exceeds AMANDA in instrumented volume.}
\label{fig:diffuse_incl_osc_gs_ai.eps}
\end{figure*}

Where do neutrinos fit into this? The assumption that the energy in neutrinos coincides with the matching energies observed in electromagnetic radiation and cosmic rays yields the level of neutrino fluxes associated with the cosmic rays shown in Fig.~\ref{neutrinosky}. In the end the neutrino flux is therefore the flux of Eq.~\ref{eq:flux}.\footnotemark\ \footnotetext{There are many corrections to the equality, from the details of the particle physics, including neutrino oscillations, to the fact that cosmic rays only reach us from within a GZK absorption length of $\sim 50$\,Mpc while neutrinos travel unimpeded from sources at all redshifts. In the end, these modifications cancel ``within a factor".} It is often referred to as the Waxman-Bahcall ''bound"\cite{wb1}. A source creating equal fluxes of cosmic rays, gamma rays and neutrinos is rather generic; see Fig.~\ref{nubeams}. The usual assumption is that cosmic rays are accelerated in a region of high magnetic fields, most likely associated with shocked particle flows driven by the gravity of a black hole. They interact with the radiation fields surrounding the black hole. The most important processes are $p + \gamma \rightarrow \Delta^+ \rightarrow \pi^0 + p$ and $p + \gamma \rightarrow \Delta^+ \rightarrow \pi^+ + n$. While the secondary protons may remain trapped in the acceleration region, roughly equal numbers of neutrons and decay products of neutral and charged pions escape. The energy escaping the source is therefore distributed among cosmic rays, gamma rays and neutrinos produced by the decay of neutrons, neutral pions and charged pions, respectively. This generic scenario accommodates the observation of equal energy in cosmic rays and electromagnetic radiation, and extends it to neutrinos. Clearly both GRB and AGN environments can accommodate this scenario although with very dissimilar black holes and radiation targets for pion production. If we take this picture seriously, our previous estimate must be corrected for the fact that the pion takes only 25\% of the energy of the secondary neutron thus changing the energy balance between cosmic rays and neutrinos and reducing their flux. In the end we estimate that the muon-neutrino flux associated with the sources of the highest energy cosmic rays is loosely confined to the range
\begin{equation}
{E_\nu}^2 dN / dE_{\nu}= 1\sim 5 \times 10^{-8}\rm\,
GeV \,cm^{-2}\, s^{-1}\, sr^{-1}
\end{equation}
depending on the cosmological evolution of the cosmic ray sources. Model calculations assuming that active galaxies or gamma-ray bursts are the actual sources of cosmic rays yield event rates similar to the generic energetics estimate presented; see Fig.~\ref{fig:diffuse_incl_osc_gs_ai.eps}.

%Fig.4

%Fig.5

The anticipated neutrino flux thus obtained has to be compared with the limit of
$7.4 \times 10^{-8}\rm\, GeV\ cm^{-2}\, s^{-1}\,sr^{-1}$ reached after the first
4 years of operation of the completed AMANDA detector in 2000--2003~\cite{hill}.  On the other hand, for conservative assumptions for the charm background and for the detector performance,  IceCube has the capability to observe a flux that is an order of magnitude below this limit with 5\,$\sigma$ significance in 3 years\cite{ice3}. The exact value of the IceCube sensitivity depends on the magnitude of the dominant high energy neutrino background from the prompt decay of atmospheric charmed particles. The level
of this background is difficult to anticipate theoretically and little accelerator data is available
in the energy and Feynman-x range of interest\cite{ingelman}.

The neutrino event rate is obtained by folding the flux predicted with the probability that
the neutrino is actually detected in a high energy neutrino telescope; only one in a million neutrinos
of TeV energy interacts and produces a muon that reaches the detector. This probability is given by the
ratio of the muon and neutrino interaction lengths in the detector medium,
$\lambda_\mu / \lambda_\nu$\cite{reviews} and therefore grows with energy. For the flux range estimated above we anticipate 100--500 detected muon neutrinos per km$^2$ per year, with the higher range close to what is already ruled out by AMANDA. In any case, the lower value represents the more realistic estimate as previously argued and the 100 events predicted will be further reduced by the realities of rejecting detector backgrounds, especially for spectra steeper than  the $E^{-2}$ assumed throughout. On the other hand, IceCube's effective area for muon neutrinos exceeds 1\,km$^2$ and equal fluxes of electron and tau neutrinos are expected\cite{ice3}. If IceCube construction remains on schedule, the instrument will accumulate 1\,km$^2$\,year of data within the next two years; the confrontation of these arguments with data is imminent, certainly on the 40 year timescale it took to develop the technology for the detectors.

We next return to galactic cosmic rays. Since the last cosmic ray conference, new observations by air Cherenkov telescopes as well as the results from an all-sky survey by the Milagro detector, are suggestively pinpointing supernova remnants as the sources. Some would say that the smoking gun is still missing and neutrinos may be the key.

\section{Cosmic Neutrinos Associated with Galactic Supernova Remnants}

Can kilometer-scale neutrino detectors observe neutrinos pointing back at the accelerators of the galactic cosmic rays? It is believed that galactic accelerators are powered by the conversion of $10^{50}$\,erg of energy into particle acceleration by diffusive shocks associated with young (1000--10,000 year old) supernova remnants expanding into the interstellar medium. The cosmic rays will interact with hydrogen atoms in the interstellar medium to produce pions that decay into photons and neutrinos. These provide us with indirect evidence for cosmic ray acceleration. The new twist here is that the eventual observation of pionic gamma rays allows for a straightforward determination of the neutrino flux.

The HESS telescope has opened a new era in astronomy by producing the first resolved images of sources in TeV gamma rays, particularly, in this context, of the supernova remnant RX J1713.7-3946\,\cite{hess}. While the resolved image of the source reveals TeV gamma ray emission from the whole supernova remnant, it shows a clear increase of the flux in the directions of known molecular clouds. This is suggestive of protons, shock-accelerated in the supernova remnant, interacting with the dense clouds to produce neutral pions that are the origin of the observed increase of the TeV photon signal. The magnitude of the photon flux is consistent with a site where protons are accelerated to energies typical of the main component of the galactic cosmic rays.  A similar extended source of TeV gamma rays tracing the density of molecular clouds has been identified near the galactic center. Protons, apparently accelerated by the remnant HESS J1745-290, diffuse through nearby molecular clouds to produce a signal of TeV gamma rays that traces the density of the clouds\,\cite{GC}. Fitting the observed spectrum by purely electromagnetic processes is challenging because the relative height of the inverse Compton and synchrotron peaks requires very low values of the $B$-field, inconsistent with those required to accelerate the electron beam to energies that can accommodate the observation of 100\,TeV photons. Nevertheless, an exclusively electromagnetic explanation of the non-thermal spectrum is not impossible, even favored by some\,\cite{Katz}. One can, for instance, partition the remnant in regions of high and low magnetic fields that are the respective sites of acceleration and inverse Compton scattering.

Supernovae associated with molecular clouds are a common feature of associations of OB stars that exist throughout the galactic plane. Although not visible to HESS, possible evidence has been accumulating for the production of cosmic rays in the Cygnus region of the galactic plane from a variety of experiments\cite{Aharonian:2002ij,Lang:2004bk,Konopelko:2006jr,Butt:2006js,Abdo:2006fq}. Most intriguing is a Milagro report of an excess of events from the Cygnus region at the $10.9~\sigma$ level\,\cite{Abdo:2006fq}. The observed flux within a $3^\circ \times 3^\circ$ window is 70\% of the Crab at the median detected energy of 12~TeV and is centered on a source previously sighted by HEGRA. Such a flux largely exceeds the one reported by the HEGRA collaboration, implying that there could be a population of unresolved TeV $\gamma$-ray sources within the Cygnus OB2 association. In fact, they report a hot spot, christened MGRO J2019+37, at right ascension $= 304.83^\circ \pm 0.14_{\rm stat} \pm 0.3_{\rm sys}$ and declination $= 36.83^\circ \pm 0.08_{\rm stat} \pm 0.25_{\rm sys}$\,\cite{Abdo:2006fq}. A fit to a circular 2-dimensional Gaussian yields a width of $0.32 \pm 0.12$ degrees, which for a distance of 1.7~kpc suggests a source radius of about 9~pc. As the brightest hotspot in the Milagro map of the Cygnus region, it represents a flux of 0.5 Crab above 12.5~TeV. A model proposed\,\cite{aongus} for MGRO J2019+37 is that of a cosmic ray beam which escapes from the OB star cluster and interacts with a molecular cloud positioned a few degrees to the southeast. Interestingly, the Tibet AS-gamma Collaboration has observed a cosmic ray anisotropy from the direction of Cygnus, which is consistent with Milagro's measurements\,\cite{Amenomori:2006bx}. As for the HESS sources, these observations suggest the production of cosmic rays as well as a variety of opportunities for neutrino production.

If the TeV gamma ray signals are indeed of pionic origin, only particle physics establishes the rate of the accompanying neutrinos. Proton-proton collisions yield two charged pions for every neutral pion, with every charged pion decaying into a muon neutrino and antineutrino (one from the pion decay, the other from the decay of the secondary muon) and a neutral pion into two photons. So, the muon neutrino flux would be equal to the photon flux were it not for a factor two reduction from oscillations. The prediction is simple; to first order there is one muon neutrino for every photon produced in the source. Because the protons transfer on average 20\% of their energy to secondary pions, and the four leptons in the charged pion decay chain $\pi \rightarrow \mu (\rightarrow e + \nu_{e} + \nu_{\mu}) + \nu_{\mu}$ take roughly equal energy, neutrinos with 0.05 of the cosmic ray energy are produced. Similarly, photons with 10\% of the proton energy originate from the decay of neutral pions. Accelerators producing cosmic rays reaching the ``knee" must produce photons with energies up to 100\,TeV and neutrinos up to half that energy. This requirement is consistent with observations of RX J1713.7-3946 and MGRO J2019+37 discussed above. They are the targets for neutrino observation of neutrino telescopes located in the Southern and Northern hemispheres, respectively.

Whereas the relation between neutrino and gamma ray fluxes is direct, the information on their spectrum is often limited. This is especially the case for the hotspot MGRO J2019+37 where we have to model the spectrum on the basis of a measurement at a single energy; the spectral slope has not been measured. Uncertainties in the calculation are associated with the propagation of the cosmic rays, with the value of the magnetic fields, and the age of the remnant. After investigating the wide parameter space of models for  MGRO J2019+37, it has been shown\cite{aongus} that the neutrino flux can nevertheless be predicted within a factor of 2 once the model flux is normalized at 12.5\,TeV to the Milagro data and a limit at GeV energy is imposed reflecting the fact that EGRET did not observe a GeV counterpart\,\cite{beacom}. The range of neutrino fluxes and event rates in IceCube are shown in Figure~\ref{fig:22events} assuming a 2.2 injected slope. The rates are within the range $2\leq dN/dt \leq 3.8$ events per year with the IceCube threshold at 50\,GeV. The range is bound by the fact that the Milagro observation strongly constrains the flux in the energy range 1--20\,TeV, where the neutrino detection probability is highest, resulting in similar predictions for dissimilar SNR characteristics.

The irreducible atmospheric background, due to neutrinos produced in the Northern atmosphere in cosmic ray showers, is calculated using the results of Ref.~\cite{agrawal}. In 15 years of operation we predict $4.5~\sigma \leq~N/\sqrt{N_{\rm atmo}}~\leq 7.7~\sigma$ and if the higher end of the predicted event rate range is realized, $5~\sigma$ is possible in 4.3 years. 

We note that the Milagro collaboration\,\cite{sinnis} has recently detected multiple additional sources besides MGRO J2019+37, most with fluxes close to 0.5 Crab. The sources with possible counterparts in the GeV range indicate a spectral index of $\sim-2.3$.  If we compute the flux of neutrinos from the Milagro sources (not including the Crab Nebula) detected with post-trial significance of greater than 5~$\sigma$ assuming a power-law index of $-2.3$, we get a total event rate in IceCube of 6.9 neutrinos/year.  If we also include the more tentative sources, the event rate increases to 11.5 neutrinos/year. In the long run, a correlation analysis of the IceCube and Milagro skymaps should make the detection of these sources likely.

%Fig.6
\begin{figure*}[t]
\centering
\includegraphics[width=5.25in]{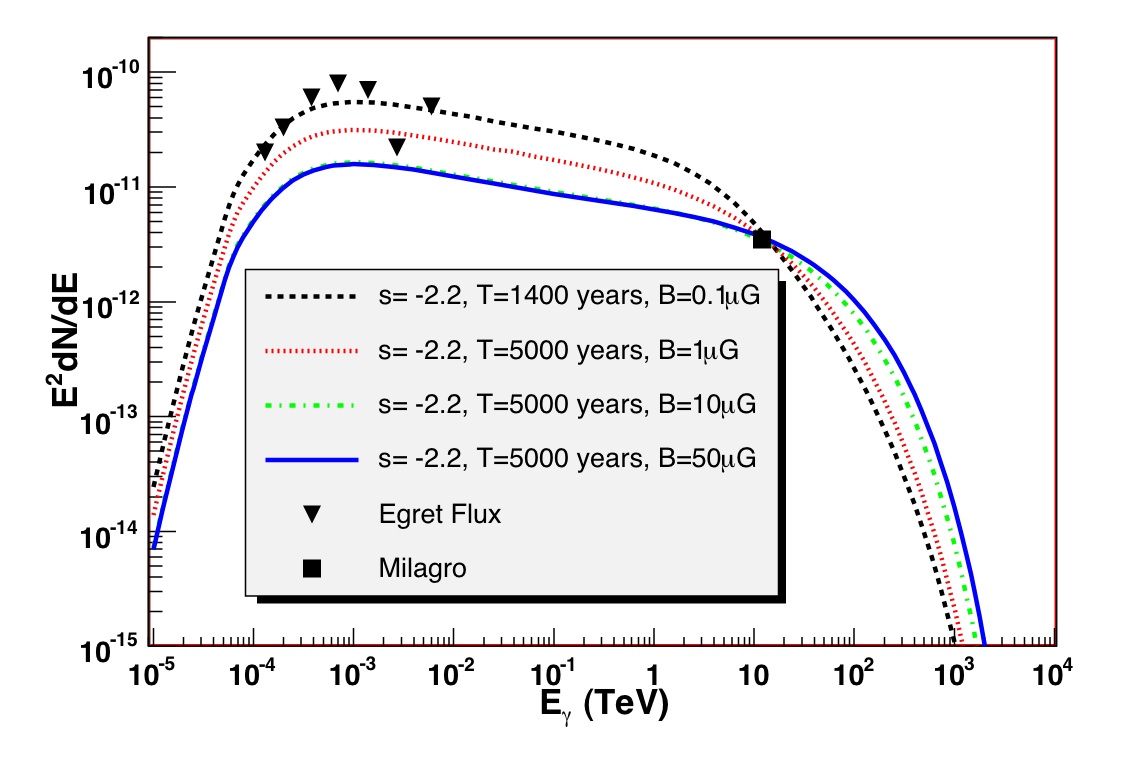}

\caption{$\gamma$-ray spectra with injection $s=-2.2$. The black dashed line is for a magnetic field of 0.1~$\mu$G and age of 1400\,years, the red dotted line is for a magnetic field of 1~$\mu$G and age of 5000\,years, the green dash-dotted line is for a magnetic field of 10~$\mu$G and age of 5000\,years, and the blue solid line is for a magnetic field of 50~$\mu$G and age of 5000\,years.
\label{fig:22spectrum}}
\end{figure*}

%Fig 7 
\begin{figure*}[t]
\centering
\includegraphics[width=4.5in]{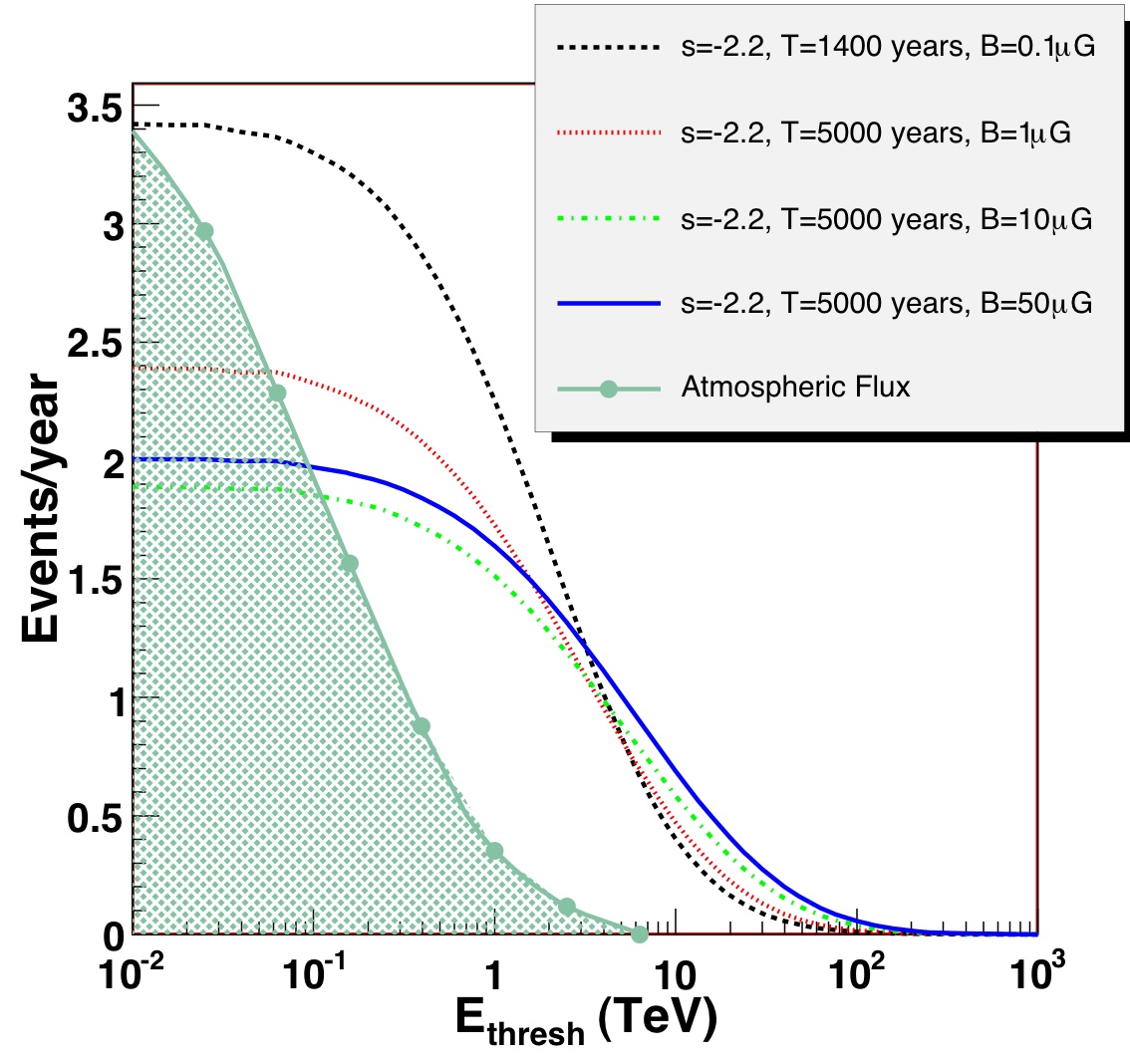}

\caption{Events due to the gamma-ray spectra with injection $\alpha=-2.2$ shown in Fig.~\ref{fig:22spectrum}.}
\label{fig:22events}
\end{figure*}

It is important to emphasize that the photon flux from the Milagro sources is consistent with the flux expected from a typical cosmic ray generating supernova remnant interacting with the interstellar medium\cite{ADV94}. In other words, the TeV flux is consistent with the energetics that are required to power the cosmic ray flux in the galaxy. Alternative candidates have been suggested for the sources of the galactic cosmic rays, for instance microquasars. Reversing the argument for supernova remnants, cosmic ray energetics requires that they should have left their imprint on the Milagro skymap and they did not. It is very suggestive that the Milagro sources are the cosmic ray accelerators.

\section{Greisen-Zatsepin-Kuzmin Neutrinos}

From our previous discussions the case for a kilometer-scale neutrino detector clearly emerges even though the predicted fluxes are hardly guaranteed. Neutrinos are guaranteed when both the accelerator beam and the pion producing target material can be identified. This is the case for GZK neutrinos produced by extragalactic cosmic rays interacting with microwave photons. The event rate is of order one per year in a kilometer-cubed detector; even kilometer-scale detectors such as IceCube are marginal in this case and unlikely to accumulate a statistically useful sample of events.  It is also important to be aware of the fact that this event rate is only determined within large ambiguities associated with the calculation of the flux and with the efficiency of the detectors at such high energies, typically EeV.

A suite of experiments have been searching for these rare events exploiting the fact that the showers initiated by EeV neutrinos emit coherent Cherenkov radiation in the $20 MHz \sim 1 GHz$ radio range\cite{radioexperiments}. EeV neutrinos can also be detected as near-horizon airshowers. The large pathlength in the atmosphere of near-horizon cosmic rays gives Auger the capabilities of an underground experiment; interesting limits have already been obtained\cite{HAS}. The possibility of exploiting this energy range with acoustic detection techniques is also under intense investigation\cite{acoustic}.

As early as 1962 Askaryan proposed using ice as a Cherenkov medium for neutrino-induced radiation of GHz wavelengths\cite{askarian}. In the early 1990s we witnessed a renewed interest in Askarian's proposal with the recognition that the relatively high neutrino energy threshold of the technique, 10 PeV or more in a reasonably scaled embedded detector in ice and even higher for other geometries, is well-matched to a number of physics goals, most notably the detection of GZK neutrinos. The RICE collaboration pioneered the technique by positioning dipole antennas in AMANDA holes at depths of several 100\,m\cite{rice}. A suite of experiments followed, GLUE and FORTE\cite{glueforte}, setting the first limits at extremely high energies exceeding $10^{20}$\,eV. More recent detectors include Lunaska\cite{ekers} and the ANITA balloon payload, which completed a prototype flight in 2004, and its first full-payload flight in early 2007.

Because of absorption by the earth, neutrinos of EeV energy predominantly produce signals in the ice near the horizon. These can refract and be detected at the ice surface or, as is the case for the ANITA experiment, by an array of 32 quad-ridged horn antennas floating in the circumpolar wind from a balloon 30\,km above Antarctica. It scans a volume of order $10^6\rm\, km^3$ of radio-transparent ice for neutrino interactions over a continent that produces very little radio noise, unlike populated regions of the world. Data from the first flight with the completed detector are subject to a blind analysis from which the results are eagerly awaited\cite{anita}.

The emission of GHz Cherenkov radiation from electromagnetic cascades is counterintuitive because, according to the Frank-Tamm formula, the power in the signal should be suppressed by a factor proportional to the frequency of $10^6$ relative to light, and is therefore unobservable. Askarian realized that this argument is wrong. At MHz$\sim$GHz frequencies the wavelength sampled exceeds the dimensions of the shower and the radiation is therefore coherent. The power coherently radiated is proportional to the square of the number of shower particles times their charge, Ne. Coherence therefore compensates for the suppression in frequency provided the number of shower particles is sufficiently large; we now know that it is detectable above the thermal background in ice provided the shower energy exceeds $\sim 10$\,PeV\cite{ZHS}. There is another subtlety here because a shower, consisting of electron-positron pairs and photons, is to a first approximation electrically neutral, $\rm Ne=0$ after summing over all particles. Askarian realized that an electromagnetic cascade develops a $\sim20$\% excess of negative charge, predominantly produced by the Compton scattering of the large number of MeV shower photons on atomic electrons. Although the complete calculation of the effect is significantly more complicated, his estimate of the charge excess was qualitatively correct\cite{ZHS}. The existence of the coherent signal was demonstrated by experiments in the SLAC beam using sand and ice targets\cite{SLAC}. The radiated power and its energy dependence measured agreed with the modern calculations.

There are additional considerations that make this technique attractive. The absorption length of radio waves in the cold ice is close to 1\,km up to 1\,GHz frequency\cite{absorption}, to be contrasted with $\sim 100$\,m for light, and their detection with antennas relatively simple. Moreover, it may be possible to perform the experiment with antenna arrays positioned at or near the surface avoiding the time consuming and expensive necessity of drilling. The array would consist of stations that can be visualized as descoped versions of the ANITA experiment, simpler yet having the possibility to make a stand-alone detection of the shower direction and energy. Already in the early 1970s Gusev and Zheleznykh\cite{gusev} proposed a surface radio array with a 10\,$\rm km^2$ footprint to detect of order 10 PeV neutrinos via antennas with grid spacing of several hundred meters.

The new Auger data presented at this meeting have cast GZK searches in a somewhat different light. The confirmation of a cutoff in the spectrum implies that the rate of super-$10^{20}$\,eV cosmic ray events is low, of order one per year. Proton astronomy, while certainly possible, will be statistically challenging. An experiment observing several GZK neutrinos per year is complementary and the neutrinos are guaranteed to point at their sources. One can indeed extract directional information from a neutrino that is produced within a GZK radius of $\sim 50$\,Mpc of a source that is at a distance of hundreds to thousands of megaparcecs. On the negative side, measurements of the depth of shower maximum ($x_{\rm max}$) suggest the possibility that a fraction of the primaries are heavy which will reduce the GZK neutrino flux. Heavy primaries photodisintegrate on the universal photon background before photoproducing the pions that are the parents of  GZK neutrinos.

In any case, we have to be realistic about the confidence with which we can anticipate the GZK flux. Despite the availability of  significantly improved data on the spectrum near and beyond the ``ankle", these can still be accommodated with a wide range of assumptions regarding the injection slope at the source, the cosmological evolution of the sources and, as already mentioned, the composition at injection. At this point scenarios yielding GZK fluxes ranging from unobservable to several per $\rm km^2\,year$ can be entertained\cite{hooper}, see Table \ref{rates} taken from reference\cite{hooper}.

%table 2
\begin{table*}
\caption{The rates of muon and shower events initiated by GZK neutrinos in a kilometer-scale neutrino telescope (such as IceCube or KM3) for a range of choices of injected spectra and chemical composition consistent with both the Auger spectrum and $X_{\rm max}$ measurements. For comparison, we also show the event rates for the case of an all-proton spectrum with injected spectral index 2.2 and $E_{\rm max} = 10^{22}$ eV. Relative to the all-proton case, models are consistent with the Auger data which range from almost no difference, to approximately two order of magnitude suppression.\label{rates}}
\medskip
\centering
\begin{tabular}{|c|c|c|c|c|}
\hline
Composition& $\alpha$& $E_{\rm max}/Z$ (eV)& Muons (km$^-2$\,yr$^-1$)& Showers (km$^3$\,yr$-1$)\\
\hline
100\% N& 1.6--1.9& 10$^{22}$& 0.20--0.0081& 0.15--0.0064\\
100\% Si& 1.6--2.0& 10$^{22}$& 0.21--0.045& 0.16--0.035\\
100\% Fe& 1.6--2.1& 10$^{22}$& 0.11--0.014& 0.085--0.012\\
100\% Fe& 1.4--1.7& 10$^{21}$& 0.019--0.0076& 0.017--0.0075\\
\hline 
50\% N, 50\% p& 1.8--2.1& 10$^{22}$& 0.23--0.013& 0.18--0.10\\
\hline
50\% Si, 50\% p& 1.6--2.1& 10$^{22}$& 0.30--0.095& 0.220--0.075\\
50\% Si, 50\% p& 1.4--1.5& 10$^{21}$& 0.059--0.051& 0.050--0.043\\
7\% Si, 93\% p& 2.0--2.2& 10$^{22}$& 0.69--0.66& 0.52--0.50\\
2\% Si, 98\% p& 1.4--1.8&10$^{21}$& 0.75--0.59& 0.60--0.47\\
\hline
50\% Fe, 50\% p& 1.6--2.1& 10$^{22}$& 0.15--0.043& 0.11--0.034\\
10\% Fe, 90\% p& 1.4--1.9& 10$^{21}$& 0.14--0.10& 0.11--0.080\\
3\% Fe, 97\% p&  2.1& 10$^{22}$& 0.68& 0.51\\
1\% Fe, 99\% p& 1.4--1.9& 10$^{21}$& 0.74--0.53& 0.59--0.43\\
\hline
100\% p (for comparison)& 2.2& 10$^{22}$& 0.76& 0.60\\
\hline
\end{tabular}
\end{table*}

Independent of the difficulty of anticipating the flux, the GZK event rate is proportional to the neutrino cross section whose calculation is challenging at these high energies where accelerator data provide little guidance. That it is poorly known has been routinely ignored in the radio business and neutrino limits from existing experiments can certainly be weakened by arguing for a less optimistic extrapolation of the neutrino cross section. In the Standard Model the neutrino cross section is simply proportional to the $q(x,Q^2) + \bar q(x,Q^2)$  quark structure function, with $Q^2 \sim M_W^2$. Sea quarks produced by gluons dominate the distribution function because, at these high energies, the relevant values of the fractional momenta of the quarks become as small as $x \sim 10^{-8}$. For the relevant $Q^2$ range, HERA data barely constrain the parton distributions to $x \sim 10^{-1}$, while collider data on inclusive jet production provide indirect evidence to $10^{-3}$. Perturbative QCD dictates a power law behavior in $x$ that allows us to extrapolate to smaller values. But what is its range of applicability? Clearly the growth of the number of constituents built into the perturbative extrapolation, and routinely assumed, cannot continue because it eventually violates the Froissart bound on the cross section. Screening of gluons will prevent this from happening. There is no consensus on the energy where this screening of the gluon constituents inside the proton, similar to that of nucleons inside nuclei, sets in, and what the mechanism is, possibly new physics such as the color glass condensate observed at RHIC. A logarithmic reduction of the cross section has been argued for on the basis of a reanalysis of the HERA data\cite{berger}.

Given the challenges in determining the flux as well as the acceptance of the detectors, one should reasonably conclude that the rate of GZK events must be determined experimentally. Ideally, the initial data of the IceCube or ANITA experiments would give us an indication. Continued data taking of IceCube and a planned ANITA flight in two years will establish GZK rates that cover a range of optimistic expectations in Table\,\ref{rates}. For instance, each experiment would detect a few events corresponding to the ``reference" flux of Engel\,\emph{et al.} which assumes a $(1+z)^3$ redshift evolution of the sources\cite{engel}. In the absence of information from the existing experiments a straightforward way to proceed is to upgrade the IceCube experiment using the optical Cherenkov technique and the infrastructure already available. The possibility has been investigated in reference \cite{halzenhooper} and the most straightforward conclusion is that significant upgrades are required, in fact on the scale of IceCube itself, to increase the event rate by ``a factor". The necessity to switch from photomultipliers to radio antennas strongly suggests itself.

The proposal is for staged deployments of radio detectors on an approximately kilometer grid expanding on the hexagonal outlay of the IceCube strings in stages of 6, 12,... radio detectors. If the event rate established with the initially deployed detectors warrants it, this array can be expanded into an instrument exceeding  $100\rm\, km^3$ in effective volume detecting hundreds of events for doing cosmic ray and particle physics. It would open up the possibility to measure the neutrino cross section at EeV energy. This is of obvious interest to particle physics; the case has been extensively illustrated in the context of TeV-scale gravity. The approach has the critical advantage that a fraction of the events will also be detected by IceCube allowing for a calibration of this novel detection method. As many as 20\% percent of the events would be hybrid events seen by both detectors provided one surrounds IceCube by an additional a single hexagon of optical strings at a distance of 500\,m as suggested in reference\cite{halzenhooper}.

Whereas calibration of these experiments would be desirable, there should no longer be any doubt that the radio technique, suggested by Askarian\cite{askarian} almost half a century ago, is robust. The signal produced by neutrinos interacting in the ice can be calculated --- it is just QED ---  and the theoretical results have been validated with accelerator experiments. There are of course additional hurdles to overcome in establishing the sensitivity of experiments that, unlike IceCube, have a threshold that is too high to allow for calibration using atmospheric neutrinos. It is important to realize that the considerable problems to establish the radio technique for detecting air showers are not relevant here, for instance the effects of the earth magnetic field which are much more difficult to quantify, are negligible.

The ARIANNA project\cite{barwick} proposes a variation on the concept for a surface radio array by positioning detector stations on the Ross Ice Shelf where the Antarctic continent is supported by water rather than rock. The experiment will capitalize on the fact that at radio frequencies the water-ice boundary below the shelf act as a mirror reflecting signals produced by neutrino signals produced in any downward direction. The concept consists of stations of cross-polarized antennas facing downward just below the snow surface positioned on a 100 by 100 grid with 300\,m spacing.

\section{Looking Forward}

While neutrino 	``telescopes" are discovery instruments with a variety of missions, the hope is that they may contribute to the resolution of the century old puzzle of the origin of cosmic rays, either by the detection of GZK neutrinos or by directly observing neutrinos from the accelerators. Between now and the next International Cosmic Ray Conference we can look forward to
\begin{itemize}
\item
The results of the first ANITA flight with a complete payload of antennas.
\item
Results from IceCube that, by operating the detector as it grows, will reach a $km^2year$ aperture.
\item
Enhanced sensitivity of the Auger experiment to neutrinos that initiate horizontal air showers.
\item
The initial design of a kilometer-scale neutrino detector in the northern hemisphere.
\item
Data as well as a wealth of ideas and initiatives on detecting the acoustic and radio signatures produced by GZK neutrinos in ice, water, salt, permafrost \dots
\end{itemize}

\section{Acknowledgments}
I have very much enjoyed the excellent hospitality of our hosts in Merida. I also thank our IceCube collaborators as well as Luis Anchordoqui, Julia Becker, Tom Gaisser, Concha Gonzalez-Garcia, Jordan Goodman, Evelyn Malkus, Teresa Montaruli, Aongus O'Murchadha and Greg Sullivan for discussions. This research was supported in part by the National Science Foundation under Grant No.~OPP-0236449, in part by the U.S.~Department of Energy under Grant No.~DE-FG02-95ER40896, and in part by the University of Wisconsin Research Committee with funds granted by the Wisconsin Alumni Research Foundation.

\end{document}